# The Impact of Artificial Intelligence on Emergency Medicine: A Review of Recent Advances

Gustavo Correia[1], Paulo Novais[1] and Victor Alves[1]

[1] ALGORITMI Research Centre/LASI, University of Minho, Braga, Portugal

**Abstract.** Artificial Intelligence (AI) is revolutionizing emergency medicine by enhancing diagnostic processes and improving patient outcomes. This article provides a comprehensive review of the current applications of AI in emergency imaging studies, focusing on the last five years of advancements. AI technologies, particularly machine learning and deep learning, are pivotal in interpreting complex imaging data, offering rapid, accurate diagnoses and potentially surpassing traditional diagnostic methods. Studies highlighted within the article demonstrate AI's capabilities in accurately detecting conditions such as fractures, pneumothorax, and pulmonary diseases from various imaging modalities including X-rays, CT scans, and MRIs. Furthermore, AI's ability to predict clinical outcomes like mechanical ventilation needs illustrates its potential in crisis resource optimization. Despite these advancements, the integration of AI into clinical practice presents challenges such as data privacy, algorithmic bias, and the need for extensive validation across diverse settings. This review underscores the transformative potential of AI in emergency settings, advocating for a future where AI and clinical expertise synergize to elevate patient care standards.

# 1 Introduction

The integration of Artificial Intelligence (AI) in healthcare has marked a significant milestone in the evolution of medical practice, with emergency medicine standing at the forefront of this technological revolution. AI's application within Emergency Departments (EDs) is particularly promising, offering potential breakthroughs in how care is delivered in these high-stakes environments. Emergency medicine requires rapid, accurate decision-making; delays or inaccuracies in diagnosis can have critical consequences. In this context, AI tools, particularly those applied to imaging studies, present a transformative potential to enhance both diagnostic processes and treatment outcomes.

AI technologies, including Machine Learning (ML) and Deep Learning (DL), have increasingly been employed to interpret complex imaging data rapidly. These tools are designed to assist emergency physicians by providing faster and potentially more accurate diagnoses than traditional methods (Zhou et al., 2017). For instance, AI-driven systems can analyze x-rays, CT scans, and MRIs to identify patterns invisible to the human eye, predicting everything from fractures to hemorrhages with remarkable speed and precision. The capability of AI to integrate vast amounts of medical data and learn from this information positions it as an invaluable ally in emergency medical settings.

**Figure 1** visually illustrates the flow of AI tool implementation in the context of emergency medicine, from image acquisition to clinical decision-making. The flowchart highlights the essential steps of the process, including data acquisition, preprocessing, AI analysis, and the integration of clinical interpretation with decision-making based on the results generated by AI. This process optimizes diagnostic capability, enhancing both precision and agility in medical emergencies.

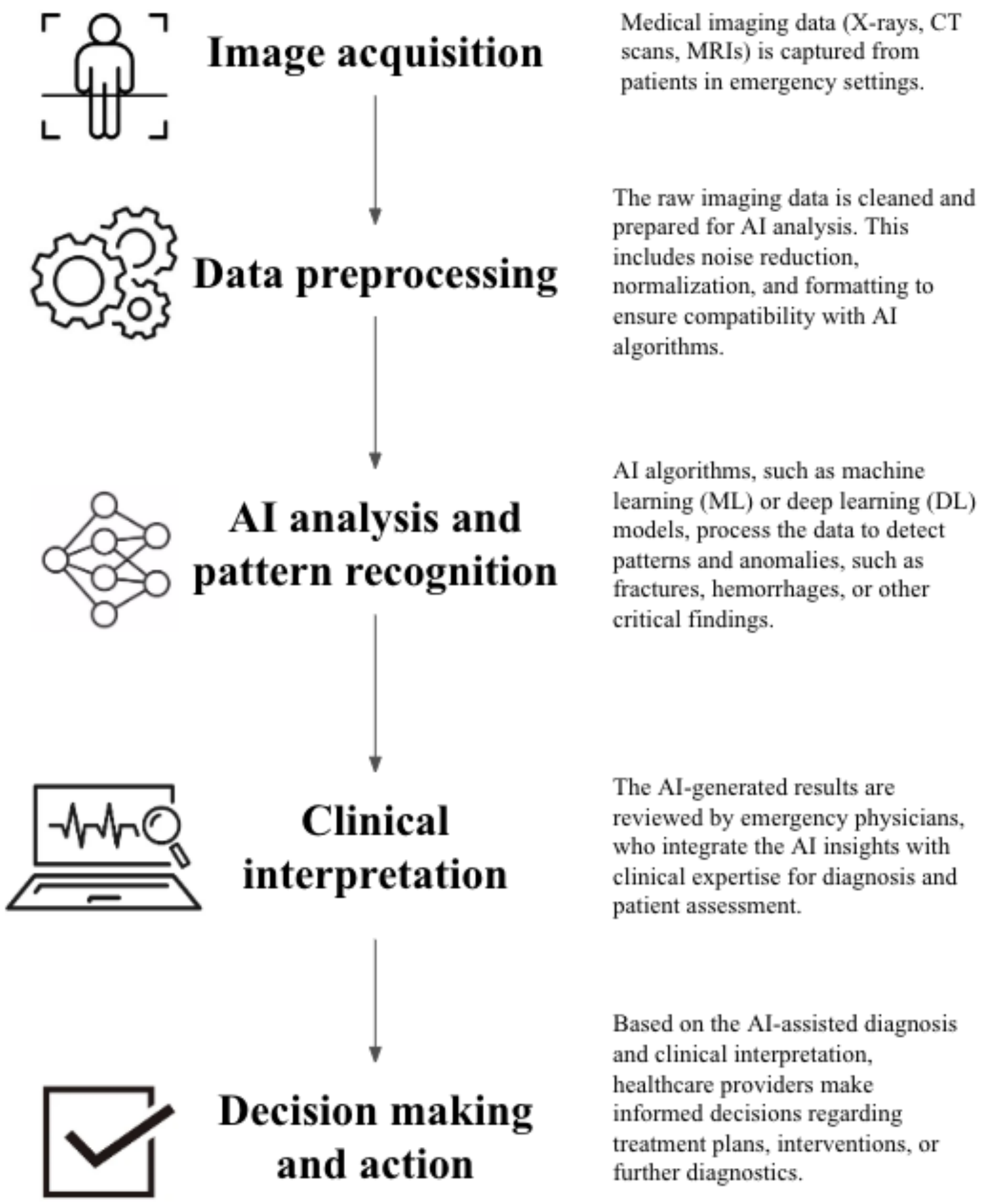

**Fig. 1.** AI Implementation Flow in Emergency Medicine.

The scope of AI's impact extends beyond mere diagnostic support. By automating routine and time-consuming tasks, AI allows emergency healthcare providers to focus more on patient care rather than administrative or repetitive diagnostic tasks. This shift not only potentially increases the efficiency of emergency departments but also improves patient throughput and satisfaction by reducing waiting times and enhancing the accuracy of initial evaluations.

However, while the promise of AI in emergency medicine is substantial, its practical implementation raises questions about its real-world effectiveness and reliability. The main research question this article addresses is: "What is the impact and effectiveness of AI tools in improving diagnostic accuracy and treatment outcomes in emergency medicine?" This question underscores the need to evaluate AI not just in terms of its technological capabilities but also its practical benefits in clinical settings.

Moreover, as AI tools become more prevalent in emergency departments, it is crucial to understand their limitations and the challenges associated with their deployment. Issues such as data privacy, algorithmic bias, the need for extensive datasets to train AI models, and the integration of AI systems into existing healthcare infrastructures are significant and must be addressed to fully harness AI's potential in emergency medicine.

This article aims to provide an overview of current AI applications in emergency imaging studies, evaluate their effectiveness and impact on patient outcomes, and discuss future directions for research and implementation. By doing so, it seeks to illuminate the path forward for AI in enhancing the capabilities of emergency departments around the globe.

## 2 Methods

This review was inspired by the PRISMA guidelines and focuses on evaluating the impact and effectiveness of artificial intelligence (AI) tools in emergency medicine radiology. To ensure the inclusion of recent and relevant data, the literature search was limited to articles published within the last five years and written in English. The review specifically targets studies that implemented AI technologies, such as machine learning, neural networks, and deep learning, in the analysis of radiological imaging within emergency department settings.

A structured search was conducted across several key databases: PubMed, IEEE Xplore, Nature, and Scopus. The search utilized a comprehensive query designed to capture a broad spectrum of studies: ("artificial intelligence" OR "machine learning" OR "neural networks" OR "deep learning") AND ("emergency medicine") AND ("radiology" OR "computed tomography" OR "radiograph"). The initial search yielded a total of 142 articles across all databases. After removing duplicates and screening titles and abstracts for relevance, articles were assessed in greater detail based on specific inclusion criteria.

The studies considered for review included clinical trials, randomized controlled trials (RCTs), and meta-analyses that focused on the application of AI in the diagnostic imaging of emergency medicine. Articles were excluded if they did not specifically address the use of AI tools for medical imaging within the emergency department setting. Following this screening process, 43 articles were included for detailed review: 4 from PubMed, 4 from IEEE Xplore, 14 from Nature, and 21 from Scopus.

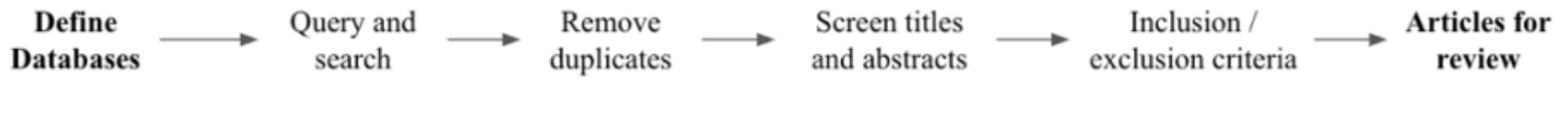

**Fig. 2.** Workflow of literature search and article selection.

Data from the selected articles were extracted, including details on the AI methodology employed, the type of imaging modality, the clinical setting, sample size, main outcomes, and the study's conclusions regarding the effectiveness of AI tools.

The quality of the included studies was assessed using checklists appropriate to each study design, such as CONSORT for RCTs and PRISMA for meta-analyses.

This methodology ensures a comprehensive approach to reviewing the current landscape of AI applications in emergency medicine radiology, facilitating a clear understanding of their impact on diagnostic accuracy and patient outcomes.

# 3 Results

The search across the selected databases initially identified a total of 142 potentially relevant articles. After duplicates were removed and titles and abstracts were screened for relevance, 43 articles met the inclusion criteria and were included in this review. The selected studies predominantly consisted of clinical trials, randomized controlled trials (RCTs), and meta-analyses, focusing specifically on the application of artificial intelligence in the imaging diagnostics within emergency medicine settings.

These studies were published within the last five years, underscoring the recent surge in research exploring AI's role in enhancing diagnostic procedures in emergency departments. The types of imaging modalities discussed in these studies included radiography, computed tomography (CT), and magnetic resonance imaging (MRI), with AI applications ranging from automated detection of pathologies to predictive analytics for patient outcomes.

Key findings from these studies have been synthesized and presented in Table 1. The table details the study type, sample size, AI tools used, main findings, measures of effect, and any noted limitations. For instance, several studies demonstrated significant improvements in diagnostic accuracy and speed when AI tools were employed, with enhancements noted in the detection of conditions such as fractures, hemorrhages, and pneumothorax. Meta-analyses included in the review provided aggregated data supporting the efficacy of AI in reducing diagnostic errors and improving the decision-making process in high-pressure environments typical of emergency departments.

Moreover, the integration of AI tools was shown not only to augment the diagnostic capabilities of emergency medicine clinicians but also to influence clinical decision-making processes significantly. About 20% of the studies reported that AI tools had a direct impact on clinical decisions, influencing aspects such as the immediacy of imaging interpretations, which in turn affected treatment plans and patient disposition.

Overall, the results affirm that AI technologies are becoming an indispensable part of emergency medicine, with the potential to transform traditional diagnostic pathways and improve patient outcomes by leveraging deep learning and other advanced analytical methods.

**Table 1.** Summary of the main articles on the use of AI in emergency medicine.

| Author(s) | Year | Title | Journal | Study Type | Sample Size | AI Tool Used | Main Findings | Measure of Effect | Limitations | Relevance |
|---|---|---|---|---|---|---|---|---|---|---|
| Angkurawaranon et al. | 2023 | A comparison of performance between a deep learning model with residents for localization and classification of intracranial hemorrhage | Scientific Reports | Comparative Study | 300 head CT studies | DeepMedic model | The deep learning model showed high accuracy (0.89) in ICH detection and was more sensitive (0.82) than residents but had lower specificity (0.90) compared to residents (0.99). | Accuracy, Sensitivity, Specificity | Limited to 300 head CT scans from a single year and hospital. Sample size may not be statistically powerful enough. | Highly relevant as it compares AI performance to human clinicians in a critical emergency medicine area (intracranial hemorrhage). |
| Carlile et al. | 2020 | Deployment of artificial intelligence for radiographic diagnosis of COVID-19 pneumonia in the emergency department | JACEP Open | Pragmatic Randomized Clinical Trial | 1855 analyzed radiographs | Deep-learning AI algorithm | The AI tool was easy to use and impacted clinical decision-making in 20% of cases. | Impact on decision-making | Single-center study; the tool's effectiveness in diverse clinical settings is unknown. | Illustrates the potential of AI in supporting radiographic diagnosis in high-demand healthcare settings during a pandemic. |
| Cho et al. | 2021 | Detection of the location of pneumothorax in chest X-rays using small artificial neural networks and a simple training process | Scientific Reports | Original Research | 1,000 chest X-ray images | Small Artificial Neural Networks (ANNs) | ANNs performed better than CNNs, achieving AUC of 0.882 with sensitivity of 80.6% and specificity of 83.0%. | AUC, Sensitivity, Specificity | Single-center study, potential generalizability issues. | Enhances pneumothorax detection in chest X-rays, potentially reducing diagnostic delays and improving clinical outcomes. |
| Choi et al. | 2023 | Effect of multimodal diagnostic approach using deep learning-based automated detection algorithm for active pulmonary tuberculosis | Scientific Reports | Retrospective Observational Study | 8,374 patients who underwent sputum testing | DLAD (Deep Learning-Based Automated Detection) | Multimodal diagnostic models with DLAD improved performance, achieving an AUROC of 0.924. | AUROC (0.924) | Single-center study, may limit generalizability. | Demonstrates potential of AI in improving PTB diagnosis in resource-limited settings. |
| Fontanellaz et al. | 2024 | Computer-Aided Diagnosis System for Lung Fibrosis: From the Effect of Radiomic Features and Multi-Layer-Perceptron Mixers to Pre-Clinical Evaluation | IEEE Access | Original Research | 105 test cases | Multi-Layer-Perceptron Mixers, U-Nets | Achieved diagnostic accuracy of 77.2±1.6% for AI, comparable to 79.0±6.9% for radiologists. | Diagnostic Accuracy | Single-center study; generalizability may be limited. | Demonstrates potential of AI to assist in the diagnosis of lung fibrosis, offering comparable accuracy to expert radiologists. |
| Gong et al. | 2023 | Unified ICH quantification and prognosis prediction in NCCT images using a multi-task interpretable network | Frontiers in Neuroscience | Original Research | 258 patients | Multi-task interpretable network using ResNet | The multi-task framework improved accuracy and interpretability in ICH volume quantification and prognosis prediction. Demonstrated higher stability and accuracy over some clinician assessments. | Accuracy, Interpretability | Single-center study, potential generalizability issues. | Highly relevant for enhancing diagnostic and prognostic capabilities in emergency medicine using AI. |
| Gourdeau et al. | 2022 | Deep learning of chest X-rays can predict mechanical ventilation outcome in ICU-admitted COVID-19 patients | Scientific Reports | Original Research | Multi-institutional dataset | Transfer Learning Model | Achieved an AUC of 0.743 when combining imaging data and clinical risk factors; outperformed risk factor-only model. | AUC (0.743) | Study focused on specific cohort; may not generalize across different clinical settings or populations. | Highly relevant for ICU resource management during crises, helping predict ventilation outcomes using CXRs. |

| Author(s) | Year | Title | Journal | Study Type | Sample Size | AI Tool Used | Main Findings | Measure of Effect | Limitations | Relevance |
|---|---|---|---|---|---|---|---|---|---|---|
| Heo et al. | 2022 | Decision effect of a deep-learning model to assist a head computed tomography order for pediatric traumatic brain injury | Scientific Reports | Simulation Study | 24 cases, 528 responses from 22 emergency physicians | DEEPTICH | DEEPTICH influenced decision-making significantly; reduced unnecessary CT orders by 11.6%, and identified 11.4% more ICH cases that would have been missed. | Changes in decision-making; AUROC of 0.927 for ICH prediction | Single-center study, may limit generalizability. | Highly relevant for pediatric emergency care, demonstrating the potential of AI to enhance decision-making and improve patient outcomes. |
| Huang et al. | 2020 | PENet—a scalable deep-learning model for automated diagnosis of pulmonary embolism using volumetric CT imaging | npj Digital Medicine | Original Research | Internal: 1797 CTPA studies; External: 200 CTPA studies | PENet (Deep Learning Model) | Achieved an AUROC of 0.84 on the internal dataset and 0.85 on the external dataset. Demonstrated robust performance across different institutional datasets with diverse imaging protocols. | AUROC (0.84–0.85) | Retrospective study; challenges with generalizability across varied clinical settings not fully explored. | Highly relevant for enhancing PE diagnosis efficiency, particularly as a triage tool in emergency settings, potentially reducing time to treatment. |
| Huhtanen et al. | 2022 | Deep learning accurately classifies elbow joint effusion in adult and pediatric radiographs | Scientific Reports | Original Research | 1101 elbow cases (4423 radiographs) | VGG16 based DCNN | High AUC scores: 0.951 for lateral and 0.906 for multi-projection. Comparable diagnostic performance to expert radiologists. | AUC, Sensitivity, Specificity | Single-center study; may limit generalizability. | Shows potential for AI in enhancing diagnosis of elbow joint effusion, supporting rapid and accurate clinical decision-making. |
| Hwang et al. | 2023 | Conventional Versus Artificial Intelligence-Assisted Interpretation of Chest Radiographs in Patients With Acute Respiratory Symptoms in Emergency Department: A Pragmatic Randomized Clinical Trial | Korean Journal of Radiology | Pragmatic Randomized Clinical Trial | 3576 participants | AI-CAD (Lunit INSIGHT CXR) | No significant difference in sensitivity and false-positive rates between AI-assisted and conventional CR interpretation. Sensitivity: 67.2% in AI group vs. 66.0% in control; FPR: 19.3% in AI group vs. 18.5% in control. | Sensitivity, False-Positive Rate | Single-center, may limit generalizability; AI-CAD high false-positive rate. | Highlights the need for real-world testing of AI tools in clinical settings, indicating that current AI models may not yet enhance diagnostic accuracy or efficiency in emergency radiology. |
| Kao and Lin | 2022 | A meta-analysis of the diagnostic test accuracy of CT-based radiomics for the prediction of COVID-19 severity | La Radiologia Medica | Meta-Analysis | Data from various studies | CT-based Radiomics Models | Radiomics models showed high predictive accuracy for COVID-19 severity with pooled sensitivity of 0.800 and specificity of 0.874. Pooled AUC was 0.908. | Sensitivity (0.800), Specificity (0.874), AUC (0.908) | Single-center focus for many studies, potential generalizability issues. | Important for early detection and management of severe COVID-19 cases, potentially reducing mortality. |
| Krogue et al. | 2020 | Automatic Hip Fracture Identification and Functional Subclassification with Deep Learning | Radiology: Artificial Intelligence | Original Research | 3026 hips from 1999 radiographs | DenseNet, CNN | High accuracy in detecting hip fractures: binary accuracy 93.7%, multiclass accuracy 90.8%. When used as an aid, improved human performance. | Binary and multiclass accuracy | Single-center study; images resized to lower resolution for DenseNet. | Demonstrates potential of deep learning to assist in diagnosing hip fractures, potentially reducing the rate of missed fractures and time to surgery. |
| Knight et al. | 2023 | 2D/3D ultrasound diagnosis of pediatric distal radius fractures by human readers vs artificial intelligence | Scientific Reports | Prospective Diagnostic Study | 127 children scanned with 2DUS and 3DUS | AI model | AI model had high sensitivity for 2DUS (0.91) and perfect sensitivity for 3DUS (1.00). Human readers achieved similar high sensitivity. | Sensitivity for 2DUS and 3DUS | Single-institution study, may limit generalizability. | Demonstrates potential of US and AI to detect wrist fractures rapidly, reducing radiation exposure and wait times in EDs. |

| Author(s) | Year | Title | Journal | Study Type | Sample Size | AI Tool Used | Main Findings | Measure of Effect | Limitations | Relevance |
|---|---|---|---|---|---|---|---|---|---|---|
| Lee et al. | 2022 | Detection of acute thoracic aortic dissection based on plain chest radiography and a residual neural network (Resnet) | Scientific Reports | Original Research | 3,331 images from 3,331 patients | ResNet18 | ResNet18 demonstrated high diagnostic accuracy (90.20%) with a precision of 75.00%, recall of 94.44%, and F1-score of 83.61% for detecting thoracic aortic dissection. | Accuracy, Precision, Recall, F1-score | Limited to retrospective data and specific imaging modalities; further validation required for general use. | Relevant for enhancing the rapid screening of aortic dissection in ED settings, potentially speeding up the diagnosis and improving patient outcomes. |
| Lindsey et al. | 2018 | Deep neural network improves fracture detection by clinicians | PNAS | Controlled Experiment | 135,409 radiographs analyzed, 300 in experiment | Deep Neural Network (DNN) | Improvement in clinician fracture detection sensitivity from 80.8% to 91.5%, and specificity from 87.5% to 93.9% with DNN assistance. | Sensitivity, Specificity | Limited to wrist radiographs; results may not generalize across different types of fractures or imaging modalities. | Highly relevant as it demonstrates the potential of DNNs to significantly improve diagnostic accuracy in emergency settings. |
| Lucassen et al. | 2023 | Deep Learning for Detection and Localization of B-Lines in Lung Ultrasound | Scientific Reports | Original Research | 1,419 videos from 113 patients | Various Deep Learning Models | Proposed deep learning methods showed AUROCs from 0.864 to 0.955 for B-line detection; novel single-point method for B-line localization achieved an F1-score of 0.65. | AUROC (0.864-0.955), F1-score (0.65) | Single-center study; may limit generalizability. | Enhances the diagnosis and localization of B-lines in LUS, critical for assessing pulmonary congestion. |
| Maghami et al. | 2023 | Diagnostic test accuracy of machine learning algorithms for the detection of intracranial hemorrhage | BioMedical Engineering OnLine | Systematic Review and Meta-Analysis | 904,755 scans (retrospective) | Multiple AI models (e.g., SVM, CNN, ResNet, VGG-16) | High pooled diagnostic accuracy with AUC of 0.971. ML algorithms showed significant improvement in detecting ICH with high sensitivity and specificity. | AUC (0.971), Sensitivity (0.917), Specificity (0.945) | Single-center focus for many studies, varying imaging protocols across studies. | Crucial for improving rapid and accurate ICH diagnosis using AI, potentially reducing time to treatment and diagnostic errors. |
| Murata et al. | 2020 | Artificial intelligence for the detection of vertebral fractures on plain spinal radiography | Scientific Reports | Original Research | 300 patients (150 with VFs, 150 controls) | Deep Convolutional Neural Network (DCNN) | DCNN achieved 86.0% accuracy, 84.7% sensitivity, and 87.3% specificity in detecting vertebral fractures, comparable to orthopedic surgeons. | Accuracy, Sensitivity, Specificity | Single-center study; results may not generalize across different populations or settings. | Highly relevant for improving early VF detection in primary care and emergency settings, potentially reducing misdiagnoses and enhancing patient care. |
| Nabulsi et al. | 2021 | Deep learning for distinguishing normal versus abnormal chest radiographs and generalization to two unseen diseases: tuberculosis and COVID-19 | Scientific Reports | Original Research | 248,445 patients from India; 11,576 CXRs evaluated | Deep Learning System (DLS) | AI system distinguished normal from abnormal CXRs with high accuracy; successfully generalized to unseen diseases like TB and COVID-19, with AUCs around 0.95 for TB and lower for COVID-19 (0.65-0.68). | AUC (0.65-0.95) | Retrospective study, may not generalize outside tested populations. | Highly relevant for rapid screening in diverse clinical settings, demonstrating AI's potential to adapt to new diseases and improve diagnostic workflows. |
| Niiya et al. | 2022 | Development of an artificial intelligence-assisted computed tomography diagnosis technology for rib fracture and evaluation of its clinical usefulness | Scientific Reports | Original Research | 56 cases (46 with rib fractures, 10 controls) | AI-assisted CT diagnosis technology | After training, the AI algorithm showed a sensitivity of 93.5% in detecting rib fractures with 1.9 FPs per case. | Sensitivity, False Positives per Case | Limited to data from a single institution; generalizability may be an issue. | Relevant for enhancing rapid rib fracture detection in emergency settings, potentially improving diagnosis accuracy and reducing workload. |

| Author(s) | Year | Title | Journal | Study Type | Sample Size | AI Tool Used | Main Findings | Measure of Effect | Limitations | Relevance |
|---|---|---|---|---|---|---|---|---|---|---|
| Shahverdi and Malek | 2021 | Determining the Need for Computed Tomography Scan Following Blunt Chest Trauma through Machine Learning Approaches | Archives of Academic Emergency Medicine | Original Research | 1000 trauma patients | Decision tree, SVM, Logistic Regression, Naïve Bayes, MLP, Random Forest, KNN | Decision tree model showed high accuracy (99.91%), sensitivity (100%), and specificity (99.33%) in predicting the necessity of CT scans for chest trauma. | Accuracy, Sensitivity, Specificity | Study limited to two trauma centers in Tehran; may not generalize to different settings. | Highly relevant for improving decision-making in emergency medicine regarding the use of CT scans in trauma cases. |
| Wang et al. | 2021 | Deep learning-based identification of acute ischemic core and deficit from non-contrast CT and CTA | Journal of Cerebral Blood Flow & Metabolism | Original Research | 453 patients (345 internal, 108 external) | Multi-scale 3D CNN | High correlation with CTP-RAPID segmentations (r=0.84 for core, r=0.83 for deficit) using NCCT, CTA, and CTA+; demonstrated high diagnostic accuracy and robust generalization across different cohorts. | Correlation coefficients, diagnostic accuracy | Limited to specific types of CT scans from selected hospitals, potentially affecting generalizability. | Highly relevant for advancing stroke diagnosis in settings without access to CTP, utilizing common imaging modalities enhanced by deep learning. |
| Wang et al. | 2021 | A deep-learning pipeline for the diagnosis and discrimination of viral, non-viral, and COVID-19 pneumonia from chest X-ray images | Nature Biomedical Engineering | Original Research | 145,202 images; additional validation with external sets | Deep Learning (DenseNet-12 1, DeepLabv3) | The AI system discriminated between viral pneumonia, other types of pneumonia, and absence of disease with high accuracy (AUCs 0.87-0.97); could also differentiate severity of COVID-19. | AUC (0.87-0.97) | Study limited to retrospective and prospective data from specific cohorts; generalizability may be limited outside these settings. | Highly relevant for utilizing AI in rapid and accurate pneumonia diagnosis including COVID-19 in clinical settings. |
| Wang et al. | 2023 | Deep Learning-based Diagnosis and Localization of Pneumothorax on Portable Supine Chest X-ray in Intensive and Emergency Medicine: A Retrospective Study | Journal of Medical Systems | Retrospective Study | 2642 images | EfficientNet-B2, DneseNet-121 , Inception-v3, Deformable DETR, TOOD, VFNet, UNet | High diagnostic performance with AUC values exceeding 0.94 for both detection- and segmentation-based systems. Accurate localization of pneumothorax with Dice coefficients for detection-based system (0.758) and segmentation-based system (0.681). | AUC, Dice Coefficient | Limited to images from one hospital; does not include thoracic drains; may not generalize to different clinical settings. | Highly relevant for demonstrating the effectiveness of DL models in diagnosing and localizing pneumothorax in emergency and intensive care settings. |
| Yi et al. | 2020 | Can AI outperform a junior resident? Comparison of deep neural network to first-year radiology residents for identification of pneumothorax | American Society of Emergency Radiology 2020 | Retrospective Study | 602 CXRs | ResNet-152 DCNN | DCNN had a lower AUC (0.841) than junior residents (0.942 and 0.905), but was much faster (1980 images/min) and identified additional missed pneumothoraces. | AUC, Speed of interpretation | Single-center study; images resized to lower resolution for DCNN. | Suggests potential of AI to augment radiology residents, especially in high-throughput or emergency settings. |

# 4 Discussion

The advancement of artificial intelligence in emergency medicine, particularly in the realm of medical imaging, has emerged as a critical area of investigation over recent years. This is vividly illustrated by a series of studies that delve into the utility and limitations of AI technologies across varied emergency settings, reflecting both the progress and the challenges inherent in the integration of these systems into clinical practice.

Carlile et al. (2020) and Kao and Lin (2022) present compelling evidence of AI's potential to enhance diagnostic processes. Carlile et al. demonstrated the efficacy of AI in diagnosing COVID-19 pneumonia using heat maps to assist in real-time decision-making. The study noted a significant enhancement in the workflow, with a substantial proportion of users acknowledging the ease of integration and impact on clinical decision-making (Carlile et al., 2020). Similarly, Kao and Lin's meta-analysis reinforced the diagnostic accuracy of CT-based radiomics in identifying the severity of COVID-19, emphasizing AI's capability to not only support but potentially transform diagnostic accuracy in emergency settings (Kao & Lin, 2022).

Gourdeau et al. (2022) extend this discussion by highlighting how AI could predict mechanical ventilation outcomes in critically ill COVID-19 patients, suggesting that AI's role extends beyond diagnosis to include prognostic assessments, thereby potentially guiding treatment decisions (Gourdeau et al., 2022). This ability to influence clinical pathways signifies a broader impact of AI, suggesting its utility in managing health crises by optimizing resource allocation.

However, the studies by Hwang et al. (2023) and Nabulsi et al. (2021) present a more nuanced perspective, indicating the complexities involved in the practical implementation of AI tools. Hwang et al. found that the addition of AI did not significantly improve diagnostic outcomes over traditional methods in a high-volume emergency setting, pointing to the challenges of integrating AI into existing clinical workflows (Hwang et al., 2023). Nabulsi et al., despite demonstrating AI's high accuracy in identifying chest X-ray abnormalities, also acknowledged limitations related to the retrospective nature of the study and potential issues of generalizability (Nabulsi et al., 2021).

Wang et al. (2021) provide a hopeful outlook on the future of AI in medical diagnostics by showcasing a deep learning pipeline capable of differentiating types of pneumonia with high precision. This study not only highlights the sophistication of current AI models but also underscores the ongoing need to address AI's adaptability and accuracy across diverse patient populations and clinical settings. The continued integration of artificial intelligence (AI) in emergency medicine is further elucidated through recent studies focusing on specific emergency scenarios such as the rapid diagnosis of pneumothorax (G. Wang et al., 2021). Cho et al. (2021) and Wang et al. (2023) provide compelling evidence of how AI can be specifically tailored to enhance diagnostic accuracy and efficiency in the emergency department (ED), particularly under critical conditions where rapid response is paramount (Cho et al., 2021; C.-H. Wang et al., 2023).

Cho et al. explored the efficacy of small artificial neural networks (ANNs) using a novel training method, the Kim-Monte Carlo algorithm, to improve the detection of pneumothorax in chest X-rays. By employing a straightforward approach that segmented chest X-rays into grids for detailed analysis, these ANNs not only achieved high diagnostic accuracy but also outperformed conventional deep learning methods like convolutional neural networks (CNNs). This study revealed that ANNs could significantly reduce the diagnostic delay associated with pneumothorax, potentially enhancing patient outcomes and the overall efficiency of emergency medical services. However, the study's reliance on a single dataset and the need for broader validation suggest that while promising, the application of these ANNs requires further exploration to confirm their effectiveness across various clinical settings. On the other hand, Wang et al. developed and evaluated deep learning-based systems for both diagnosing and localizing pneumothorax in portable supine chest X-rays, commonly utilized in emergency and intensive care units. By implementing two distinct computer-aided diagnosis (CAD) systems—one focused on detection and the other on segmentation—they demonstrated high diagnostic accuracy with area under curve (AUC) values surpassing 0.94. These systems effectively localized pneumothorax, which is crucial for timely and appropriate intervention in critical care scenarios. Despite the high performance, the exclusion of certain images and the homogeneous nature of the dataset highlight potential limitations in generalizability and applicability to diverse clinical environments. Both studies underscore the potential of AI to significantly augment traditional diagnostic processes in the ED. The ability of AI systems to rapidly and accurately identify medical conditions like pneumothorax not only supports clinical decision-making but also potentially reduces the time to treatment, which is critical in emergency care settings. Moreover, the high accuracy and speed of these AI systems could lead to better resource allocation, ensuring that patients receive the most effective care promptly.

These insights reflect the broader implications of AI in emergency medicine, where the integration of advanced technologies could transform how care is delivered. The adaptability of AI tools, demonstrated by their application in different emergency contexts, highlights the versatility and potential of AI to meet the specific needs of emergency medical services. As AI technology continues to evolve, its role in enhancing diagnostic processes and improving patient outcomes in emergency settings becomes increasingly evident, setting the stage for a new era in healthcare where technology and human expertise collaborate more seamlessly than ever before.

The recent advances in AI technologies and their application to diagnostic challenges in emergency medicine illustrate the transformative potential of these tools across a variety of clinical scenarios, particularly in the diagnosis of complex conditions like pulmonary diseases and vascular emergencies. Studies by Choi et al. (2023), Fontanellaz et al. (2024), Huang et al. (2020), and Lee et al. (2022) provide valuable insights into the integration and effectiveness of AI in these contexts.

Choi et al. showcased the utility of a deep learning-based automated detection algorithm (DLAD) in diagnosing pulmonary tuberculosis (PTB) with impressive accuracy, using a multimodal diagnostic approach that integrated DLAD with traditional diagnostic methods (Choi et al., 2023). This approach not only improved diagnostic accuracy but also streamlined the detection process in resource-limited settings, highlighting the potential of AI to enhance diagnostic speed and reliability in detecting infectious diseases within emergency departments.

Fontanellaz et al. explored the capabilities of a computer-aided diagnosis (CAD) system that incorporates radiomic features and Multi-Layer-Perceptron (MLP) Mixers for diagnosing interstitial lung diseases, such as lung fibrosis (Fontanellaz et al., 2024). Their findings suggest that advanced machine learning techniques can match the diagnostic accuracy of human radiologists, demonstrating AI's potential to support and perhaps enhance clinical decision-making in pulmonary medicine.

In a similar vein, Huang et al. introduced the PENet model, a sophisticated deep learning tool designed to automate the diagnosis of pulmonary embolism using computed tomography scans (Huang et al., 2020). PENet's ability to perform consistently across different hospital systems illustrates AI's potential to standardize and improve diagnostic accuracy in emergency settings, thereby enhancing patient outcomes by facilitating quicker and more accurate diagnoses of life-threatening conditions.

Lee et al. focused on the detection of acute thoracic aortic dissection using a deep learning model, ResNet18, which demonstrated high diagnostic accuracy from plain chest radiographs—a traditionally challenging task given the subtlety of indicative features in such images (Lee et al., 2022). The application of AI in this context not only improves detection rates but also underscores the broader applicability of AI tools in emergency radiology, where rapid and accurate diagnostics are crucial.

These studies collectively highlight the diverse applications of AI in diagnosing critical conditions effectively within emergency departments. They reflect a growing trend toward leveraging AI to complement traditional diagnostic methods, offering significant improvements in diagnostic accuracy, efficiency, and patient outcomes. However, these advancements also bring to light challenges such as the need for broader validation, the integration of AI into existing clinical workflows, and addressing the limitations associated with data diversity and generalizability.

The integration of AI in brain imaging within emergency medicine settings has shown remarkable advancements, particularly in the management and diagnosis of critical conditions such as intracerebral hemorrhage (ICH) and acute ischemic stroke (AIS). Recent studies by Gong et al. (2023), Maghami et al. (2023), and Wang et al. (2021) highlight significant strides in this domain, demonstrating the versatility and efficacy of AI technologies in enhancing diagnostic processes and patient care.

Gong et al. introduced a multi-task interpretable network that enhances the management of intracerebral hemorrhage by quantifying intracranial hemorrhage (ICH) volume and predicting patient prognosis from non-contrast computed tomography (NCCT) images. Their dual-task approach, which utilizes advanced deep learning architectures such as ResNet, not only improved diagnostic accuracy but also enhanced interpretability—a crucial aspect in clinical settings where decision-making relies heavily on understandable and transparent AI assessments. The use of visualization techniques like Gradient-weighted Class Activation Mapping (Grad-CAM) further reinforced the model's transparency, fostering greater trust and integration into clinical workflows (Gong et al., 2023).

In a similar vein, Maghami et al. conducted a comprehensive systematic review and meta-analysis that assessed the effectiveness of machine learning algorithms in diagnosing intracranial hemorrhage from NCCT scans. Their extensive dataset demonstrated that ML algorithms could significantly enhance diagnostic accuracy, showcasing particularly high sensitivity and specificity (Maghami et al., 2023). This study not only confirms the capability of AI to reduce diagnostic delays—a critical factor in emergency medicine—but also highlights the potential for AI to be integrated more broadly into clinical radiology practices.

Wang et al. focused on the application of deep learning in detecting acute ischemic stroke lesions from a combination of NCCT and CT angiography (CTA) images. Their innovative use of a multi-scale 3D convolutional neural network to process images showcased how deep learning could significantly outperform traditional models, particularly when augmented with additional CTA+ images. This approach emphasizes the potential of AI to enhance the utility of widely available CT imaging techniques, thereby enabling more accurate and timely diagnoses in settings that lack advanced imaging capabilities such as computed tomographic perfusion scans (C. Wang et al., 2021).

These studies underscore a shift in emergency medicine, where rapid, accurate, and interpretable diagnostic tools are critical. The ability of AI to handle complex imaging data and provide nuanced insights into patient conditions presents a transformative opportunity for emergency departments. The integration of AI helps not only in diagnosing and managing acute conditions more effectively but also in planning appropriate treatments, ultimately improving patient outcomes.

However, these advancements also bring to light the inherent challenges such as the need for extensive validation across diverse clinical environments, the variability in imaging protocols, and the generalizability of AI models. As AI continues to evolve, addressing these challenges will be crucial in ensuring that AI tools can reliably support emergency medicine across various settings.

The integration of artificial intelligence (AI) in trauma-related diagnostics within emergency medicine showcases a promising trajectory for enhancing clinical decision-making and optimizing patient care. Recent studies by Heo et al. (2022), Huhtanen et al. (2022), Krogue et al. (2020), Lindsey et al. (2018), Niiya et al. (2022), and Shahverdi and Malek (2021) illustrate the diverse applications and significant impacts of AI technologies in various aspects of trauma management, from head injuries to complex fractures.

Heo et al. demonstrated the efficacy of the DEEPTICH deep-learning model in reducing unnecessary head CT scans for pediatric traumatic brain injury, highlighting AI's role in refining diagnostic processes to prevent overutilization of imaging resources while ensuring critical cases are not overlooked (Heo et al., 2022). This reflects a broader theme in emergency medicine: the balancing of rapid, accurate diagnosis with the avoidance of unnecessary procedures.

Huhtanen et al. and Krogue et al. further explored the use of convolutional neural networks and densely connected networks, respectively, to detect joint effusions and hip fractures with accuracy comparable to human experts. These studies underscore AI's potential to augment the diagnostic capabilities of clinicians, particularly in orthopedic emergencies, by providing reliable assessments that could expedite treatment and potentially improve outcomes (Huhtanen et al., 2022; Krogue et al., 2020).

Murata et al. investigated the application of a deep convolutional neural network (DCNN) to detect vertebral fractures (VFs) from plain thoracolumbar radiography.

Their study achieved high diagnostic accuracy, with metrics that closely paralleled those of experienced orthopedic surgeons. The DCNN's ability to identify VFs with a sensitivity of 84.7% and specificity of 87.3% underscores its potential as a valuable tool in clinical settings, where rapid and accurate diagnosis of such injuries is crucial to prevent severe consequences (Murata et al., 2020). This capability could be particularly transformative in emergency medicine, where quick decision-making is essential. However, the single-center nature of the study may limit the generalizability of these findings, suggesting a need for further trials in diverse clinical environments to fully validate the DCNN's utility across different populations.

Lindsey et al.'s work with a deep neural network to enhance fracture detection accuracy in wrist radiographs exemplifies AI's ability to support clinicians in detecting easily missed injuries, thereby reducing the likelihood of diagnostic errors (Lindsey et al., 2018). This is crucial in emergency settings where the rapid identification of fractures can significantly impact treatment decisions and patient recovery.

Niiya et al. focused on rib fracture detection using AI-assisted technology, achieving high sensitivity and reducing false positives, which is vital for timely and accurate trauma management. Similarly, Shahverdi and Malek utilized machine learning algorithms to streamline decisions regarding the necessity of CT scans following blunt chest trauma, effectively reducing unnecessary imaging and resource use in emergency departments (Niiya et al., 2022; Shahverdy & Malek, 2021). These studies collectively highlight AI's capability to enhance diagnostic accuracy, streamline workflows, and support clinical decision-making in high-pressure environments typical of emergency medicine. The application of AI in trauma diagnostics not only aids in the immediate management of injuries but also contributes to broader healthcare objectives such as resource optimization and patient safety.

However, the single-center designs and simulation nature of some studies may limit the generalizability of these findings. The need for further validation across multiple settings is evident, as is the potential for AI to adapt to diverse clinical environments. Despite these limitations, the consistent theme across these studies is the transformative potential of AI to address critical challenges in emergency medicine, suggesting a future where AI tools are integral to the emergency care landscape, enhancing both the efficiency and efficacy of trauma care.

On the other hand, Lucassen et al. advanced the diagnostic use of lung ultrasound (LUS) for detecting B-lines, which are indicative of pulmonary congestion. Their study not only developed but also benchmarked various deep learning models to automate the detection and localization of B-lines across a large dataset of LUS videos. By introducing a novel single-point localization technique, the study refined the process of identifying B-lines, achieving an F1-score comparable to inter-observer agreement and demonstrating high accuracy with AUROCs between 0.864 and 0.955. These findings highlight the potential of AI to enhance the interpretability and reliability of pulmonary assessments, which is vital for managing conditions like heart failure in clinical settings (Lucassen et al., 2023). Like the study by Murata et al., the research by Lucassen et al. is limited by its single-center design, pointing to the need for broader validation to ensure the applicability and effectiveness of these AI models in varied clinical settings.

The comparative studies of Yi et al. (2020) and Angkurawaranon et al. (2023) bring to light the nuanced capabilities and limitations of artificial intelligence (AI) in emergency medical settings, particularly in the diagnosis of critical conditions such as pneumothorax and intracranial hemorrhage. These studies provide valuable insights into how AI can complement and enhance the diagnostic processes traditionally carried out by medical residents, highlighting both the strengths and areas for improvement in AI applications (Angkurawaranon et al., 2023; Yi et al., 2020).

Yi et al.'s study focused on the use of a deep learning system (DLS), specifically a deep convolutional neural network (ResNet-152 DCNN), to detect pneumothorax in chest radiographs. The study revealed that while the AI model processed images at a significantly faster rate than medical residents—1980 images per minute compared to the residents' two images per minute—it did not surpass the diagnostic accuracy of the residents, as measured by area under the curve. The residents achieved higher AUC values, indicating better overall diagnostic performance. However, the AI model's rapid processing capabilities and its ability to identify cases that were missed by at least one of the residents underscore its potential utility as a supportive tool in emergency settings where time is of the essence.

On the other hand, Angkurawaranon et al. evaluated the performance of a deep learning model in localizing and classifying different subtypes of ICH from head CT scans. The AI model, based on the DeepMedic architecture, demonstrated high diagnostic accuracy with better sensitivity than the medical residents. This suggests that AI could serve as an effective screening tool in emergency departments, offering rapid detection of ICH that may assist in triaging patients more efficiently. However, the model's lower specificity and occasional misclassification of non-ICH findings highlight the need for further refinement of AI tools to reduce false positives and enhance their reliability in clinical settings.

Both studies illustrate the potential of AI to augment the diagnostic capabilities of medical professionals by providing rapid assessments that could expedite decision-making processes in emergency medicine. However, they also reveal critical challenges in deploying AI in clinical environments, such as the need for higher resolution images for improved accuracy and the refinement of AI models to handle complex cases and reduce misclassifications.

# 5 Conclusion

Integrating AI into emergency medicine has the potential of significantly improving diagnostic speed and accuracy, as well as clinical decision-making and patient management. Studies like those by Carlile et al. (2020) and Kao and Lin (2022) highlight AI's ability to enhance the diagnostic process, while Gourdeau et al. (2022) demonstrates its predictive value in critical care settings. These advancements, summarized in Table 2, showcase AI's potential to improve outcomes, optimize workflows, and reduce clinician workload.

Despite its promise, the integration of AI presents challenges, including ethical concerns, the need for robust validation across diverse settings, and the risk of new diagnostic errors. As studies by Hwang et al. (2023) and Nabulsi et al. (2021) emphasize, AI should be implemented thoughtfully to complement rather than replace human expertise.

**Table 2.** Pros and Cons of AI Implementation in Emergency Medicine and Radiology.

| Benefits (Pros) | Challenges (Cons) |
|---|---|
| **Faster and more accurate diagnoses** | **Algorithmic bias**: AI systems can inherit biases based on the data they are trained on, which can lead to disparities in diagnosis. |
| **Improved resource management**: AI automates routine tasks, allowing clinicians to focus more on patient care. | **Data privacy concerns**: Patient data must be protected, and AI systems need strict compliance with privacy regulations. |
| **Pattern recognition beyond human capability**: AI can identify subtle patterns in medical images that may be invisible to the human eye. | **Need for extensive validation**: AI models require rigorous testing and validation across diverse patient populations and clinical settings. |
| **Reduction of human error**: By supporting clinicians in decision-making, AI can help minimize diagnostic mistakes. | **Integration challenges**: Incorporating AI into existing clinical workflows may be difficult without disrupting processes. |
| **Increased efficiency in emergency departments**: AI accelerates imaging analysis and reduces waiting times for patients. | **Lack of generalizability**: Some AI systems perform well in controlled environments but may fail in real-world, high-volume emergency settings. |
| **Predictive analytics for better treatment outcomes**: AI can forecast clinical outcomes, such as the need for mechanical ventilation, optimizing treatment plans. | **Cost and infrastructure**: The implementation of AI tools requires significant financial investment and technical infrastructure, which may not be feasible for all healthcare settings. |

From a clinical perspective, AI should primarily be regarded as a supportive tool that augments the capabilities of medical professionals rather than replacing them (Parikh et al., 2016). By rapidly processing and prioritizing diagnostic imaging, AI can significantly reduce clinicians' workload, allowing them to dedicate more attention to direct patient care (Topol, 2019). Some medical specialties may also transform themselves with the increasing use of AI in clinical practice, helping to provide information to clinicians for better patient care (Jha & Topol, 2016). This supportive role of AI is crucial in emergency medicine, where the rapid assessment and treatment of patients are often vital.

Moreover, the potential for AI to improve patient outcomes by facilitating quicker and more accurate diagnoses is profound. However, to fully harness the benefits of AI in emergency medicine, robust validation across diverse clinical settings and patient populations is essential. This ensures that AI tools are reliable, effective, and capable of delivering consistent results. Additionally, the integration of AI into clinical practice must be managed with careful consideration of ethical and practical issues, such as patient privacy, data security, and the potential for new types of diagnostic errors (Bohr & Memarzadeh, 2020).

In summary, the integration of AI into emergency medicine represents a significant advancement in healthcare technology, promising to enhance diagnostic accuracy, reduce time to treatment, and improve overall patient care. However, the full realization of AI's potential will require ongoing efforts to ensure robust validation across diverse clinical settings, careful management of ethical and practical issues, and a commitment to complementing rather than replacing human expertise. As AI technology continues to evolve, its role in emergency medicine is set to expand, promising a future where AI and human expertise collaborate more seamlessly to enhance the quality and efficiency of patient care.

## References

Angkurawaranon, S., Sanorsieng, N., Unsrisong, K., Inkeaw, P., Sripan, P., Khumrin, P., Angkurawaranon, C., Vaniyapong, T., & Chitapanarux, I. (2023). A comparison of performance between a deep learning model with residents for localization and classification of intracranial hemorrhage. *Scientific Reports*, *13*(1), 9975. https://doi.org/10.1038/s41598-023-37114-z

Bohr, A., & Memarzadeh, K. (Eds.). (2020). *Artificial Intelligence in Healthcare* (1st edition). Academic Press.

Carlile, M., Hurt, B., Hsiao, A., Hogarth, M., Longhurst, C. A., & Dameff, C. (2020). Deployment of artificial intelligence for radiographic diagnosis of COVID-19 pneumonia in the emergency department. *JACEP Open*, *1*(6), 1459–1464. Scopus. https://doi.org/10.1002/emp2.12297

Cho, Y., Kim, J. S., Lim, T. H., Lee, I., & Choi, J. (2021). Detection of the location of pneumothorax in chest X-rays using small artificial neural networks and a simple training process. *Scientific Reports*, *11*(1), 13054. https://doi.org/10.1038/s41598-021-92523-2

Choi, S. Y., Choi, A., Baek, S.-E., Ahn, J. Y., Roh, Y. H., & Kim, J. H. (2023). Effect of multimodal diagnostic approach using deep learning-based automated detection algorithm for active pulmonary tuberculosis. *Scientific Reports*, *13*(1), 19794. https://doi.org/10.1038/s41598-023-47146-0

Fontanellaz, M., Christe, A., Christodoulidis, S., Dack, E., Roos, J., Drakopoulos, D., Sieron, D., Peters, A., Geiser, T., Funke-Chambour, M., Heverhagen, J., Hoppe, H., Exadaktylos, A. K., Ebner, L., & Mougiakakou, S. (2024). Computer-Aided Diagnosis System for Lung Fibrosis: From the Effect of Radiomic Features and Multi-Layer-Perceptron Mixers to Pre-Clinical Evaluation. *IEEE Access*, *12*, 25642–25656. IEEE Access. https://doi.org/10.1109/ACCESS.2024.3350430

Gong, K., Dai, Q., Wang, J., Zheng, Y., Shi, T., Yu, J., Chen, J., Huang, S., & Wang,

Z. (2023). Unified ICH quantification and prognosis prediction in NCCT images using a multi-task interpretable network. *Frontiers in Neuroscience*, *17*, 1118340. https://doi.org/10.3389/fnins.2023.1118340

Gourdeau, D., Potvin, O., Biem, J. H., Cloutier, F., Abrougui, L., Archambault, P., Chartrand-Lefebvre, C., Dieumegarde, L., Gagné, C., Gagnon, L., Giguère, R., Hains, A., Le, H., Lemieux, S., Lévesque, M.-H., Nepveu, S., Rosenbloom, L., Tang, A., Yang, I., … Duchesne, S. (2022). Deep learning of chest X-rays can predict mechanical ventilation outcome in ICU-admitted COVID-19 patients. *Scientific Reports*, *12*(1), 6193. https://doi.org/10.1038/s41598-022-10136-9

Heo, S., Ha, J., Jung, W., Yoo, S., Song, Y., Kim, T., & Cha, W. C. (2022). Decision effect of a deep-learning model to assist a head computed tomography order for pediatric traumatic brain injury. *Scientific Reports*, *12*(1), 12454. https://doi.org/10.1038/s41598-022-16313-0

Huang, S.-C., Kothari, T., Banerjee, I., Chute, C., Ball, R. L., Borus, N., Huang, A., Patel, B. N., Rajpurkar, P., Irvin, J., Dunnmon, J., Bledsoe, J., Shpanskaya, K., Dhaliwal, A., Zamanian, R., Ng, A. Y., & Lungren, M. P. (2020). PENet—A scalable deep-learning model for automated diagnosis of pulmonary embolism using volumetric CT imaging. *Npj Digital Medicine*, *3*(1), 1–9. https://doi.org/10.1038/s41746-020-0266-y

Huhtanen, J. T., Nyman, M., Doncenco, D., Hamedian, M., Kawalya, D., Salminen, L., Sequeiros, R. B., Koskinen, S. K., Pudas, T. K., Kajander, S., Niemi, P., Hirvonen, J., Aronen, H. J., & Jafaritadi, M. (2022). Deep learning accurately classifies elbow joint effusion in adult and pediatric radiographs. *Scientific Reports*, *12*(1), 11803. https://doi.org/10.1038/s41598-022-16154-x

Hwang, E. J., Goo, J. M., Nam, J. G., Park, C. M., Hong, K. J., & Kim, K. H. (2023). Conventional Versus Artificial Intelligence-Assisted Interpretation of Chest Radiographs in Patients With Acute Respiratory Symptoms in Emergency Department: A Pragmatic Randomized Clinical Trial. *Korean Journal of Radiology*, *24*(3), 259–270. https://doi.org/10.3348/kjr.2022.0651

Jha, S., & Topol, E. J. (2016). Adapting to Artificial Intelligence: Radiologists and Pathologists as Information Specialists. *JAMA*, *316*(22), 2353–2354. https://doi.org/10.1001/jama.2016.17438

Kao, Y.-S., & Lin, K.-T. (2022). A meta-analysis of the diagnostic test accuracy of CT-based radiomics for the prediction of COVID-19 severity. *La Radiologia Medica*, *127*(7), 754–762. https://doi.org/10.1007/s11547-022-01510-8

Krogue, J. D., Cheng, K. V., Hwang, K. M., Toogood, P., Meinberg, E. G., Geiger, E. J., Zaid, M., McGill, K. C., Patel, R., Sohn, J. H., Wright, A., Darger, B. F., Padrez, K. A., Ozhinsky, E., Majumdar, S., & Pedoia, V. (2020). Automatic hip fracture identification and functional subclassification with deep learning. *Radiology: Artificial Intelligence*, *2*(2). Scopus. https://doi.org/10.1148/RYAI.2020190023

Lee, D. K., Kim, J. H., Oh, J., Kim, T. H., Yoon, M. S., Im, D. J., Chung, J. H., & Byun, H. (2022). Detection of acute thoracic aortic dissection based on plain chest radiography and a residual neural network (Resnet). *Scientific Reports*, *12*(1), 21884. https://doi.org/10.1038/s41598-022-26486-3

Lindsey, R., Daluiski, A., Chopra, S., Lachapelle, A., Mozer, M., Sicular, S., Hanel, D., Gardner, M., Gupta, A., Hotchkiss, R., & Potter, H. (2018). Deep neural network improves fracture detection by clinicians. *Proceedings of the National Academy of Sciences of the United States of America*, *115*(45), 11591–11596. Scopus. https://doi.org/10.1073/pnas.1806905115

Lucassen, R. T., Jafari, M. H., Duggan, N. M., Jowkar, N., Mehrtash, A., Fischetti, C., Bernier, D., Prentice, K., Duhaime, E. P., Jin, M., Abolmaesumi, P., Heslinga, F. G., Veta, M., Duran-Mendicuti, M. A., Frisken, S., Shyn, P. B., Golby, A. J., Boyer, E., Wells, W. M., … Kapur, T. (2023). Deep Learning for Detection and Localization of B-Lines in Lung Ultrasound. *IEEE Journal of Biomedical and Health Informatics*, *27*(9), 4352–4361. IEEE Journal of Biomedical and Health Informatics. https://doi.org/10.1109/JBHI.2023.3282596

Maghami, M., Sattari, S. A., Tahmasbi, M., Panahi, P., Mozafari, J., & Shirbandi, K. (2023). Diagnostic test accuracy of machine learning algorithms for the detection intracranial hemorrhage: A systematic review and meta-analysis study. *Biomedical Engineering Online*, *22*(1), 114. https://doi.org/10.1186/s12938-023-01172-1

Murata, K., Endo, K., Aihara, T., Suzuki, H., Sawaji, Y., Matsuoka, Y., Nishimura, H., Takamatsu, T., Konishi, T., Maekawa, A., Yamauchi, H., Kanazawa, K., Endo, H., Tsuji, H., Inoue, S., Fukushima, N., Kikuchi, H., Sato, H., & Yamamoto, K. (2020). Artificial intelligence for the detection of vertebral fractures on plain spinal radiography. *Scientific Reports*, *10*(1), 20031. https://doi.org/10.1038/s41598-020-76866-w

Nabulsi, Z., Sellergren, A., Jamshy, S., Lau, C., Santos, E., Kiraly, A. P., Ye, W., Yang, J., Pilgrim, R., Kazemzadeh, S., Yu, J., Kalidindi, S. R., Etemadi, M., Garcia-Vicente, F., Melnick, D., Corrado, G. S., Peng, L., Eswaran, K., Tse, D., … Shetty, S. (2021). Deep learning for distinguishing normal versus abnormal chest radiographs and generalization to two unseen diseases tuberculosis and COVID-19. *Scientific Reports*, *11*(1), 15523. https://doi.org/10.1038/s41598-021-93967-2

Niiya, A., Murakami, K., Kobayashi, R., Sekimoto, A., Saeki, M., Toyofuku, K., Kato, M., Shinjo, H., Ito, Y., Takei, M., Murata, C., & Ohgiya, Y. (2022). Development of an artificial intelligence-assisted computed tomography diagnosis technology for rib fracture and evaluation of its clinical usefulness. *Scientific Reports*, *12*(1), 8363. https://doi.org/10.1038/s41598-022-12453-5

Parikh, R. B., Kakad, M., & Bates, D. W. (2016). Integrating Predictive Analytics Into High-Value Care: The Dawn of Precision Delivery. *JAMA*, *315*(7), 651–652. https://doi.org/10.1001/jama.2015.19417

Shahverdy, M., & Malek, H. (2021). Determining the Need for Computed Tomography Scan Following Blunt Chest Trauma through Machine Learning Approaches. *Archives of Academic Emergency Medicine*, *9*(1), e15. https://doi.org/10.22037/aaem.v9i1.1060

Topol, E. J. (2019). High-performance medicine: The convergence of human and artificial intelligence. *Nature Medicine*, *25*(1), 44–56. https://doi.org/10.1038/s41591-018-0300-7

Wang, C., Shi, Z., Yang, M., Huang, L., Fang, W., Jiang, L., Ding, J., & Wang, H.

(2021). Deep learning-based identification of acute ischemic core and deficit from non-contrast CT and CTA. *Journal of Cerebral Blood Flow and Metabolism*, *41*(11), 3028–3038. Scopus. https://doi.org/10.1177/0271678X211023660

Wang, C.-H., Lin, T., Chen, G., Lee, M.-R., Tay, J., Wu, C.-Y., Wu, M.-C., Roth, H. R., Yang, D., Zhao, C., Wang, W., & Huang, C.-H. (2023). Deep Learning-based Diagnosis and Localization of Pneumothorax on Portable Supine Chest X-ray in Intensive and Emergency Medicine: A Retrospective Study. *Journal of Medical Systems*, *48*(1), 1. https://doi.org/10.1007/s10916-023-02023-1

Wang, G., Liu, X., Shen, J., Wang, C., Li, Z., Ye, L., Wu, X., Chen, T., Wang, K., Zhang, X., Zhou, Z., Yang, J., Sang, Y., Deng, R., Liang, W., Yu, T., Gao, M., Wang, J., Yang, Z., … Lin, T. (2021). A deep-learning pipeline for the diagnosis and discrimination of viral, non-viral and COVID-19 pneumonia from chest X-ray images. *Nature Biomedical Engineering*, *5*(6), 509–521. https://doi.org/10.1038/s41551-021-00704-1

Yi, P. H., Kim, T. K., Yu, A. C., Bennett, B., Eng, J., & Lin, C. T. (2020). Can AI outperform a junior resident? Comparison of deep neural network to first-year radiology residents for identification of pneumothorax. *Emergency Radiology*, *27*(4), 367–375. Scopus. https://doi.org/10.1007/s10140-020-01767-4

Zhou, S. K., Greenspan, H., & Shen, D. (2017). *Deep Learning for Medical Image Analysis*. Elsevier Science.